\documentclass[aps,pra,twocolumn,superscriptaddress,showpacs,fleqn,floatfix]{revtex4-1}
\usepackage{graphicx,amssymb}
\usepackage[fleqn]{amsmath}
\usepackage{amsfonts}
\usepackage{dcolumn}
\usepackage{bm}
\usepackage{braket}
\usepackage{multirow} 
\usepackage{MnSymbol}

\usepackage{CJKutf8}

\graphicspath{{images/}}

\renewcommand{\vec}[1]{\mbox{\boldmath $#1$}}

\newcommand{\lmax}
{{ \ell }_{ \text{max} }}

\newcommand{\kmax}
{{k}_{ \text{max} }}

\begin{document}
\begin{CJK*}{UTF8}{gbsn}

\date{\today}

\title{Resonant spectra of quadrupolar anions}

\author{K. Fossez}
\affiliation{NSCL/FRIB Laboratory,
Michigan State University, East Lansing, Michigan  48824, USA}

\author{Xingze Mao (毛兴泽)}
\affiliation{NSCL/FRIB Laboratory,
Michigan State University, East Lansing, Michigan  48824, USA}

\author{W. Nazarewicz}
\affiliation{Department of Physics and Astronomy and FRIB Laboratory,
Michigan State University, East Lansing, Michigan  48824, USA}
\affiliation{Institute of Theoretical Physics, Faculty of Physics, 
University of Warsaw, 02-093 Warsaw, Poland}

\author{N. Michel}

\affiliation{NSCL/FRIB Laboratory,
Michigan State University, East Lansing, Michigan  48824, USA}

\affiliation{Grand Acc\'el\'erateur National d'Ions Lourds (GANIL), CEA/DSM - CNRS/IN2P3, BP 55027, F-14076 Caen Cedex, France}

\author{W. R. Garrett}
\affiliation{Department of Physics and Astronomy,
University of Tennessee, Knoxville, Tennessee 37996, USA}

\author{M. P{\l}oszajczak}
\affiliation{Grand Acc\'el\'erateur National d'Ions Lourds (GANIL), CEA/DSM - CNRS/IN2P3,
BP 55027, F-14076 Caen Cedex, France}

\begin{abstract}
In quadrupole-bound anions, an extra electron is attached at a sufficiently large quadrupole moment of a neutral molecule, which is lacking a permanent dipole moment. The nature of the bound states and low-lying resonances of such anions is of interest for understanding the threshold behavior of open quantum systems in general. In this work, we investigate the properties of quadrupolar anions as extreme halo systems, the formation of rotational bands, and the transition from a subcritical to supercritical electric quadrupole moment. We solve the electron-plus-molecule problem using a non-adiabatic coupled-channel formalism by employing the Berggren ensemble, which explicitly contains bound states, narrow resonances, and the scattering continuum. We demonstrate that binding energies and radii of quadrupolar anions strictly follow the scaling laws for two-body halo systems. Contrary to the case of dipolar anions, ground-state band of  quadrupolar anions smoothly extend into the continuum, and many rotational bands could be identified above the detachment threshold. We study the evolution of a bound state of an anion as to dives into the continuum  at a critical  quadrupole moment and we show that the associated critical exponent is consistent with the second-order phase transition. Everything considered, quadrupolar anions represent a perfect laboratory for the studies of marginally bound open quantum systems.
\end{abstract}

\maketitle
\end{CJK*}

\section{Introduction}

Multipolar anions form a unique class of molecular open quantum systems  \cite{desfrancois96_114,compton01_b60,jordan03_352,simons08_1079},
whose properties arise from the competition between the short-ranged electrostatic multipolar potential, 
nonadiabatic coupling of electronic motion to molecular rotation, and a strong coupling to the one-electron continuum.
They thus represent prototypical extreme halos \cite{riisager92_615,jensen04_233,mitroy05_235,knoop09_1087,hammer10_1093,ferlaino10_580,frederico12_372,stipanovic14_1190}, and laboratories for similar systems, such as deformed halo nuclei that are governed by short-ranged potentials, non-adiabatic rotation, and a strong coupling to the one- or two-nucleon continuum.
Moreover,  general studies  of resonances 
reveal a plethora of phenomena present in open quantum systems, 
such as exceptional points \cite{zirnbauer83_1022,okolowicz09_911}, 
 superradiance  \cite{auerbach11_879}, 
 near-threshold clustering  \cite{oertzen06_1017,freer07_1018,okolowicz13_241,okolowicz12_998}, 
and  resonance trapping \cite{kleinwachter85_1488}.
Consequently, unique characteristics of multipolar molecules in the landscape of open quantum systems call for detailed  studies  of their resonant spectra.

The striking case of dipolar anions has been extensively studied \cite{desfrancois96_114,compton01_b60,jordan03_352,simons08_1079}, 
using effective potentials methods \cite{desfrancois95_205,desfrancois96_114,abdoul98_108,abdoul02_1321,garrett70_118,garrett71_106,garrett82_104,ard09_122,fossez13_552} 
as well as ab initio approaches \cite{jordan77_232,gutsev98_1097,adamowicz89_346,smith99_355,clary99_356,kalcher00_560,skurski02_575,peterson02_123}. 
In this case, the attractive potential ${ 1/r^2 }$ is singular and thus can support an infinite number of bound states \cite{landau77_b96}
for a  dipole moment that is greater than a certain critical value.
This raised a question whether dipole-bound anions could be a realization of a quantum anomaly \cite{camblong01_845}. However, when considering the rotational  
motion of the system, the answer to this question is negative  \cite{coon02_846,bawin04_844,bawin07_39}.

The value of the critical moment $\mu_c$ required to bind an extra electron has been first determined \cite{turner77_1037} by Fermi and Teller \cite{fermi47_110} 
for a point-dipole (${ \mu_c = 1.625\,\text{D} }$) and then generalized to an extended dipole with an infinite moment of inertia \cite{levy67_115}.
However, high resolution electron photodetachment experiments \cite{lykke84_309,marks88_366,andersen91_1106,brinkman93_208,mullin93_209,ard09_122} 
suggested a greater critical moment, which appeared to be consistent with nonadiabatic calculations (${ \mu_c \sim 2.5\,\text{D} }$) 
including the rotational degrees of freedom of the anion \cite{garrett70_118,garrett71_111,crawford71_1323,garrett80_105,garrett80_120,garrett81_302,garrett82_104,clary88_345,fossez13_552}.
In this case, dipolar anions support only few bound states and the value of $\mu_c$ depends on the moment of inertia.

Moreover, rotational states in dipolar anions were expected to be strongly affected by the shallowness of the molecular potential 
and the nonadiabatic coupling of the electronic and molecular rotational motions \cite{herrick84_1248,clary88_345,clary89_303,brinkman93_208,ard09_122,garrett10_117}.
The strong coupling of the attached electron to the continuum \cite{estrada84_294,clark84_1095,fabrikant85_1104,mccartney90_1105,martorell08_1098,nicolaides10_b131,nicolaides12_b132} 
renders the picture even more complex,
with the existence of low-energy sharp resonances \cite{edwards12_351,wong74_308,lykke84_309,rohr75_283,rohr76_284,rohr78_1426,ziesel75_1427} in various systems, 
and the modification of the Wigner's law \cite{wigner48_591,baz58_1429,baz61_1430} for the dipolar field \cite{malley65_1081,garrett81_302,sadeghpour00_297,chernov05_305}, 
first observed in hydrogen atoms \cite{seaton61_1428,gailitis63_1425} and then extended to different power-law potentials \cite{malley65_1081,sadeghpour00_297}.

In a previous study, 
using a pseudopotential method and the Berggren expansion technique \cite{fossez15_1028}, we showed 
that in hydrogen cyanide anions ${ \text{HCN}^-}$
the competition between continuum effects and rotations leads to a transition from the subthreshold strong-coupling regime, 
where the external electron is in a spatially extended bound halo state that follows the rotational motion of the molecule, 
to the weak-coupling regime, 
where the electron and molecular motions are largely decoupled.

The quadrupolar anions case seems more straightforward.
The potential ${ 1/r^3 }$ is singular for any attractive value of the quadrupole moment and its asymptotic solutions are known analytically
for a finite moment of inertia of the molecule.
However, it has been an experimental challenge to find systems where the excess electron is bound only due to the electric quadrupole moment \cite{simons08_1079,klahn98_752}.
While the attachment of an electron by a pure quadrupolar field has been proposed theoretically already in 1979 \cite{jordan79_330} for beryllium oxide anions (BeO)$_2^-$, 
the first experimental evidence of a quadrupole-bound anion has been obtained only in 2004 \cite{desfrancois04_199}, 
for the trans-succinonitrile (${ \text{NC}-\text{CH}_{2}-\text{CH}_{2}-\text{CN} }$) molecule, 
whose {gauche} conformer is actually dipole-bound.
Even octupole-bound anions have been reported experimentally \cite{gutowski00_1337} as early as in 2000.
Recently, state-of-the-art ab initio calculations on several quadrupole-bound anion candidates \cite{sommerfeld04_1573,sommerfeld14_1187} 
concluded  that the quadrupole binding is  much weaker  than dipole binding as 
the electron-molecule potential is not dominated by one component.

One of the reason is that the attractive ${ 1/r^3 }$ potential can attach an  electron in very localized states \cite{abdoul98_108,abdoul02_1321}, 
near the neutral polar molecule.
It is thus difficult if not impossible in realistic conditions, 
to disentangle whether the electron's binding energy  comes solely from the long-range quadrupolar field \cite{desfrancois98_1336,gutsev99_1324,gutsev98_1317}.
As a case in point, the (BeO)$_2^-$ anion has a rhombic neutral form with a zero dipole-moment and a large quadrupole moment, 
and is quadrupole-bound in its ground state (g.s.) according to ab initio calculations \cite{gutowski99_1186}, 
but a critical quadrupole moment cannot be defined as for dipolar anions.

A controversial example remains the carbon disulfide 
CS$_2^-$
quadrupole-bound candidate.
On the one hand, it can be argued that its quadrupole moment, 
$-$2.67466\,$e {a}_{0}^{2}$, is insufficient to attach an electron \cite{gutowski99_1186}, 
but, on the other hand, it has a simple linear geometry in its neutral form,
and ab initio calculations \cite{gutsev98_1317} have shown that CS$_2^-$ can exist in an excited linear configuration that
is stable towards autodetachment by about  0.0012\,Ry, an energy compatible with binding energies in polar anions.
Experimentally, the situation is also difficult to interpret \cite{allan03_1319}.
Indeed, Rydberg electron transfer data \cite{kalamarides88_151,harth89_1326,carman90_1328,barsotti02_1318,suess03_1185} 
show a characteristic sharp peak in the Rydberg effective principal number 
$ n^*$$ \sim$17  dependence 
for the formation of CS$_2^-$ that undergo electric-field-induced detachment.
This is usually understood as a signature of dipole- or quadrupole-bound states  \cite{compton96_152,crawford71_1323}.

While CS$_2^-$ has a positive electron affinity of $\sim$0.05\,Ry \cite{oakes86_1434,klaus80_1490,gutsev98_1317},
it does not easily attach a free electron directly to form a long-lived anion due to rapid autodetachment \cite{wang86_1355}, 
and  a stabilization process is required for its formation \cite{harth89_1326}.
Moreover, the g.s. of CS$_2^-$ is bent \cite{benz85_1327,carman90_1328,harth89_1326,oakes86_1434} with an angle of about ${ 132^\circ }$.
The predicted linear excited state of CS$_2^-$  thus appears as a ``doorway state" to the g.s. \cite{compton96_152,popple95_1340}.
This is also suggested by the fact that the ${ n^* \sim 17 }$ energy is close to the bending vibrational energy 
of the CS$_2^-$ g.s. \cite{carman90_1328,harth89_1326,allan03_1319}.
Whether the linear excited state of CS$_2^-$ is a pure quadrupole-bound state or not, cannot be answered using simple models.
However, simple models allow to shed light on particular aspects of problems, 
such as, for example, the role of the rotational motion in the critical binding of an electron on an electric dipole.

We propose to investigate general properties of linear quadrupole-bound and unbound anions, 
using an electron-plus-molecule model and by taking into account the particle continuum.
In this picture, the linear core is represented as a triad of point charges \cite{prasad89_135,prasad91_136} separated by a distance $s$, 
with two possible configurations: oblate ${ ( -q , 2q , -q ) }$ and prolate  ${ ( q , -2q , q ) }$ with ${ q > 0 }$.
Considering the cylindrical symmetry along the molecule axis (${ z }$-axis), 
and according to the Buckinkham convention \cite{buckingham59_1424}, 
the quadrupole moment is given by ${ Q_{zz}^\pm = \pm 2 q s^2 }$, 
where the sign of $Q_{zz}$ is given by  the sign of the extremal charges in the triads.

For this simple geometry of the system and in the adiabatic limit, i.e., for an infinite moment of inertia of the neutral molecule, 
it is possible to calculate very precisely the positive and negative critical electric quadrupole moments of the core
required to attach an excess electron \cite{ferron04_130} in a ${ J^{ \pi } = 0^+ }$ state.
In Ref.~\cite{ferron04_130}, using the finite-scaling method \cite{privman90_b204} recently introduced in atomic physics \cite{neirotti97_1466,serra98_1464,kais00_1465}, 
the critical quadrupole moments have been expressed as a function of the scaled parameter ${ q_s = qs }$, 
so that ${ Q_{zz}^\pm = \pm 2 q_s s }$.
The critical values for the scaled parameter have been found to be: ${ q_{s,c}^+ = 3.98251 \, (e a_0) }$ and ${ q_{s,c}^- = 1.46970 \, (e a_0) }$, 
for prolate and oblate critical quadrupole moments, respectively, 
and are consistent with numerical calculations \cite{pupyshev04_161}.
The quality of these estimates comes from the analyticity of the scaling property of the critical quadrupole moments.
For that reason, we shall refer to these results as analytical in the following, even if they were arrived at numerically.

This paper is organized as follows.
The model Hamiltonian is presented in Sec.~\ref{model_and_methods}, 
as well as the coupled channel formulation of the Schr\"odinger equation and 
the methods used to solve it. 
The results are discussed in Sec.~\ref{results}.
We  benchmark our numerical calculations by comparing them to the analytical  values of the critical electric quadrupole moments. Thereafter follows the general discussion of
quadrupolar anions as extreme halo systems, and we analyze 
the properties of the g.s. band in the continuum.
Finally, we discuss resonant spectra, with an emphasis on quasi-degenerate states. We also study the evolution of  resonant states with the electric quadrupole moment.
The conclusion and outlook are contained in Sec.~\ref{conclusion}.

\section{Model and Methods}\label{model_and_methods}
	
	\subsection{Hamiltonian}

The schematic description of quadrupolar anions, in terms of a neutral molecular core plus an attached electron, is partially justified by the scale separation  between binding energies of valence electron of the neutral molecule, and the  energy attachment of the extra electron. However,  microscopic studies have shown that for the quadrupolar potential the attached electron in the g.s. configuration is still rather close to the ``core". In this study, however, we are interested in the low-energy states of quadrupolar anions, and in particular their resonances. For these very extended states the scale separation argument applies well \cite{bertulani02_869,bedaque03_1085}.

The rotational degrees of freedom of the molecular core are included 
	within the particle-plus-rotor model \cite{bohr98_b27} 
	in a non-adiabatic manner as described in Refs.~\cite{garrett10_117,fossez13_552,fossez15_1028}.
	Moreover, since the attached electron is assumed to be rather far from the core, the spin-orbit interaction is neglected.
	In this picture, the model Hamiltonian can be written as:
	\begin{equation}
		H = \frac{ \vec{p}_{e}^{2} }{ 2 {m}_{e} } + \frac{ \vec{j}_{r}^{2} }{ 2 I } + V
		\label{eq_Hamiltonian}
	\end{equation}
	where ${ \vec{p}_{e} }$ is the linear momentum of the attached electron, 
	$m_e$ -- its mass, 
	and $I$ is the moment of inertia of the molecule.
	The total angular momentum is thus  ${ \vec{J} = \vec{\ell} + \vec{j}_{r} }$, 
	with ${ \vec{\ell} }$ being the orbital angular momentum of the electron, 
	and ${ \vec{j}_{r} }$ the  molecular angular momentum.
	The pseudopotential $V$ that describes the interaction 
	between the core and the electron \cite{garrett08_131} is expressed through a multipole expansion:
	\begin{equation}
		V ( r , \theta ) = \sum_{ \lambda } {V}_{ \lambda } (r) {P}_{ \lambda } ( \cos\theta  ),
		\label{eq_quad_pot_multipole_expan}
	\end{equation}
	where the radial part ${ {V}_{ \lambda } (r) }$ is the electrostatic potential of a the linear charge distribution ${ ( \pm q , \mp 2q , \pm q ) }$:
	\begin{align}\label{eq_quad_pot_rad_def}
		{V}_{\lambda} (r) =  \frac{e}{4\pi {\varepsilon}_{0}} \frac{ {Q}^{\pm} }{ {s}^{2} } 
		\begin{cases}
		   \frac{1}{r_>} - \frac{1}{r} & \text{for } \lambda = 0  \\
	\left( \frac{r_<}{r_>} \right)^\lambda \frac{1}{r_>}  & \text{for } \lambda =2,4,6\dots 
			\end{cases}
	\end{align}
	with ${ {r}_{>} = \max ( r , s ) }$ and ${ {r}_{<} = \min ( r , s ) }$.

	\subsection{Coupled channel equations}

	The Schr\"odinger equation can be conveniently expressed 
	in the coupled-channel (CC) formalism, 
	where the total wave function for a given total angular momentum ${ {J}^{ \pi } }$ can be written as:
	\begin{equation}
		{ \Psi }^{ {J}^{ \pi } } (r) = \sum_{c} {u}_{c}^{ {J}^{ \pi } } (r) { \Theta }_{c}^{ {J}^{ \pi } },
		\label{eq_CC_wfs}
	\end{equation}
	where the index ${ c }$ labels the channels ${ ( \ell , {j}_{r} ) }$, 
	and ${ {u}_{c}^{ {J}^{ \pi } } (r) }$ and ${ { \Theta }_{c}^{ {J}^{ \pi } } }$ 
	are the radial and angular channel wave functions, respectively.
	Since the Hamiltonian is rotationally invariant, 
	the wave function is independent of the total angular momentum projection ${ {M}_{J} }$.

	The CC equations are obtained by inserting the ansatz \eqref{eq_CC_wfs} in the Schr\"odinger equation:
	\begin{align}
		&\left[ \frac{ {d}^{2} }{ d {r}^{2} } - \frac{ \ell ( \ell + 1 ) }{ {r}^{2} } - \frac{ {j}_{r} ( {j}_{r} + 1 ) }{I} + {E}^{ {J}^{ \pi } } \right] {u}_{c}^{ {J}^{ \pi } } (r) \nonumber \\
		&= \sum_{ c' } {V}_{ c c' }^{ {J}^{ \pi } } (r) {u}_{ c' }^{ {J}^{ \pi } } (r)
		\label{eq_CC_eqs}
	\end{align}
	where  ${ {V}_{ c , c' }^{ {J}^{ \pi } } }$ is the channel-channel coupling potential \cite{fossez13_552}.

	\subsection{Berggren expansion method}

	To solve the CC equations, we apply two methods.
	The first is the conventional Direct Integration Method (DIM), 
	described in Ref.~\cite{fossez13_552}. In DIM,
	one integrates the CC equations from a given starting energy. This method
	 gives very precise results when considering a limited number of channels, and bound states or fairly narrow resonances.
	The second method is the Berggren Expansion Method (BEM), 
	described in Refs.~\cite{fossez13_552,fossez15_1028}, 
	which may give results slightly less precise than the DIM, if the latter applies, 
	but much better results for a large number of channels and for broad resonances.
	Moreover, since this technique is based on a diagonalization approach,
	it does not require any starting energy to converge and yields the full spectrum.

	In the BEM, each channel wave function in Eq.~\eqref{eq_CC_wfs} is expanded in a single particle (s.p.) basis, 
	the so-called Berggren basis \cite{berggren68_32}, 
	originally developed for configuration-interaction calculations in nuclear physics \cite{michel09_2}.
	The Berggren basis is a generalization of the Newton basis \cite{newton82_b6} in the complex plane;
	it explicitly contains bound states, decaying resonances, and scattering continuum.
	The construction of the Berggren basis for each partial wave $c$
	is done as follows. In the first step,
the discrete resonant (Gamow) solutions ${ { \phi }_{c} ( {k}_{i} ) }$ 
			of a given spherical one-body generating potential
		are calculated 	assuming the outgoing boundary conditions.
In the next step, the bound states (${ {k}_{i} }$ imaginary) and decaying resonances (${ {k}_{i} = { \alpha }_{i} - i { \beta }_{i} }$, ${ { \alpha }_{i}, { \beta }_{i} > 0 }$) 
			that are relevant for the description of a physical system 
			are selected and surrounded by a contour ${ \mathcal{L}_{c}^{+} }$ of complex-energy scattering states ${ { \phi }_{c} (k) }$  to ensure the completeness.

 The completeness relation for the resulting Berggren basis corresponding to a channel $c$ is:
\begin{equation}
		\sum_{i} \ket{ { \phi }_{c} ( {k}_{i} ) } \bra{ \tilde{ \phi }_{c} ( {k}_{i} ) } + \int_{ \mathcal{L}_{c}^{+} } dk \, \ket{ { \phi }_{c} (k) } \bra{ \tilde{ \phi }_{c} (k) } = \hat{1}
		\label{eq_Berggren_basis_c}
	\end{equation}
	where the contour ${ \mathcal{L}_{c}^{+} }$ starts at zero, surrounds the selected resonances and extends to ${ k \to +\infty }$.
	The tilde symbols indicate time-reversal.
	One may notice that there is some freedom when it comes to the choice of the Berggen basis: 
	the form of the generating potential; 
	the selection of the discrete resonant states entering the completeness relation; and 
	the form of the contour ${ \mathcal{L}_{c}^{+} }$.

	In the present study, 
	the Berggren basis for each partial wave is generated 
	using the diagonal elements ${ {V}_{cc} (r) }$ of the channel-channel coupling potential; such a choice
	 improves the convergence of calculations.
	Because of the Cauchy's integral theorem, the precise form of the contour ${ \mathcal{L}_{c}^{+} }$ is unimportant, 
	provided that all the selected discrete states lie between the contour and the real axis in the momentum plane.

	The normalization of bound states is standard, 
	while for decaying resonant states this is accomplished by means of the exterior complex scaling \cite{dykhne61_1041,gyarmati71_38,simon79_436}.
	The scattering states are normalized to the Dirac delta.
	%
	In practical applications, 
	the integral along the contour ${ \mathcal{L}_{c}^{+} }$ in Eq.~\eqref{eq_Berggren_basis_c} 
	is discretized using the Gauss-Legendre quadrature, and 
	the selected scattering states are renormalized by the quadrature weights.
	The normalization of discretized scattering states reduces in practice to the Kronecker delta normalization.
	In the calculations presented in this study, 
	the shape of the contour has been defined through three segments: 
	the first segment connecting the origin and the point ${ {k}_{ \text{peak} } = k_r - i k_i }$ with ${ k_r, k_i > 0 }$;
	the second segment connecting points ${ {k}_{ \text{peak} } }$ and ${ {k}_{ \text{middle} } }$ (real);
	and the third segment lying on the real axis  between ${ {k}_{ \text{middle} } }$ and ${ {k}_{ \text{max} } }$.
	The momentum cutoff ${ {k}_{ \text{max} } }$ has to be sufficiently large 
	to ensure the completeness of the Berggren basis.

	Since the Berggren basis explicitly contains bound states, resonances and scattering states, 
	it is ideally suited for the description of very diffuse systems, 
	such as halos or Rydberg states, 
	and also for unstable resonant states.
	While the DIM is of limited applicability when the initial energy required to ensure the convergence  
	has to be chosen very close to the exact value, 
	the BEM may also suffer from a related problem.
	Indeed, the discrete states entering the Berggren basis 
	are obtained by integrating the Schr\"odinger equation with ${V}_{cc} (r)$, 
	which is a process that requires a choice of starting energy.
	In many situations, harmonic oscillator expansion of the potential 
	provides a starting point that is good enough to ensure the converge of the integration method, 
	but for very weakly bound states or long-range potentials, this may fail.
	For that reason, a different approach, less sensitive to the initial conditions, has been proposed.

	The idea is to use the fact that the quality of the integration method 
	with respect to the starting energy, 
	is deteriorating  faster than convergence speed of the eigenvalue ${ {E}_{f} < 0 }$.
	Thus, for a potential ${ W ( \eta ) }$ that has a bound state with ${ E \to -\infty }$ when ${ \eta \to +\infty }$, 
	it is always possible to find a starting energy ${ {E}_{0} }$ so that the integration ${ \mathcal{I} ( {E}_{0} , { \eta }_{0} ) }$ will always converge 
	for a sufficiently large value of ${ { \eta }_{0} > 0 }$.
	Once such a point has been found, 
	it is possible to make the integration to converge to the physical eigenenergy ${ {E}_{f} < 0 }$ 
	at the physical value of ${ { \eta }_{f} > 0 }$ that defines the actual potential.

	Indeed, the initial eigenenergy ${ {E}_{0} ( { \eta }_{0} ) }$ 
	can be used as a starting energy to obtain ${ {E}_{1} ( { \eta }_{1} ) = {E}_{1} ( { \eta }_{0} + \Delta{ \eta } ( {E}_{0} ) ) }$ 
	with ${ { \eta }_{0} > { \eta }_{1} \geq { \eta }_{f} }$.
	The same operation can be repeated using ${ {E}_{1} ( { \eta }_{1} ) }$ 
	as a starting energy to get ${ {E}_{2} ( { \eta }_{2} ) = {E}_{2} ( { \eta }_{0} + \Delta{ \eta } ( {E}_{0} ) + \Delta{ \eta } ( {E}_{1} ) ) }$, 
	with ${ { \eta }_{1} > { \eta }_{2} \geq { \eta }_{f} }$.
	After ${ N+1 }$ iterations, one gets:
	\begin{equation}
		\sum_{n = 0}^{N} \Delta{ \eta } ( {E}_{n} ) = { \eta }_{f} - { \eta }_{0}
		\label{eq_quad_mom_algo_sum_cond}
	\end{equation}
	In order to minimize the number of iterations, the partition of ${ { \eta }_{f} - { \eta }_{0} }$ 
	can be chosen to exploit the sensitivity of the direct integration with respect to the starting energy, 
	which is increasing as ${ E \to {0}^{-} }$.
	Thus the steps ${ \Delta{ \eta } ( {E}_{n} ) }$ 
	must be decreasing as ${ E \to {0}^{-} }$, 
	to both (i) preserve the stability of the integration at each step 
	and  (ii) minimizing the number of steps by considering bigger steps for larger values of ${ E < 0 }$.

	To perform the partition of ${ { \eta }_{f} - { \eta }_{0} }$, 
	any series ${ {u}_{n} }$ with ${ {u}_{N} = 0 }$ and ${ {u}_{n} > {u}_{n+1} }$, 
	and which preserves the stability of the integration, would suffice.
	If by  ${ {U}_{N} }$ one denotes the sum of  ${ {u}_{n} }$, 
	then the steps are defined by:
	\begin{equation}
		\Delta{Q} ( {E}_{n} ) = ( {Q}_{f} - {Q}_{0} ) \frac{ {u}_{n} }{ {U}_{N} }
		\label{eq_steps_quad_mom_def_series}
	\end{equation}
	In our case, ${ {u}_{n} = 1/(n+1) - 1/(N+1) }$
	has provided a good compromise.
Such an  improved iterative procedure for bound  states turned out to be
helpful for evaluating the critical value of the parameter ${ \eta = {Q}_{zz,c}^{ \pm } }$ of the quadrupolar potential, as in this case  extraordinary accuracy and stability are required. 

	\subsection{Identification of resonances}

	In the Berggren basis, the Hamiltonian matrix becomes complex symmetric
	even if the Hamiltonian itself  is Hermitian.
	This has a direct practical consequence, 
	since the diagonalization of the Hamiltonian matrix gives a set of eigenstates 
	that contain the resonant spectrum (bound states and  resonances) embedded in the discretized complex-energy scattering continuum.
	Because we are interested in resonant states,  an identification procedure has to be used to identify them.


	In the absence of poles in the Berggren basis, 
	the overlap method \cite{michel02_8,michel03_10} usually applied in nuclear physics, 
	based on the assumption that continuum states play a perturbative role, cannot be applied.
	In this case, one may rely on another property of physical solutions.
	Indeed, resonant states given by the diagonalization in the full space 
	are a priori independent of the precise form of the contour ${ \mathcal{L}_{c}^{+} }$.
	The contour-independence of resonant solutions has been used to identify dipolar anion resonances in Ref.~\cite{fossez15_1028}.
	In the present study, we also utilize this technique.
	To this end, we take two contours ${ \mathcal{L}_{0}^{+} }$ and ${ \mathcal{L}_{1}^{+} }$, which differ by the imaginary part of ${ {k}_{ \text{peak} } }$ and are discretized using the same number of points.  
	While scattering solutions obtained with these contours are shifted along the  imaginary axis, the resonant states are fairly insensitive as 
	 the precise shape of the contour does not impact decaying solutions.
	For the identification of very weakly bound states and low-lying resonances, 
	that are only given as a superposition of complex-energy scattering states in the BEM, 
	the method based on the concept of contour independence has been essential.

\section{Results}
	\label{results}

	\subsection{Critical quadrupole moments}\label{critical_Q}

		In order to benchmark the DIM and BEM as applied to quadrupolar anions, our adiabatic-limit results are compared with the analytical results of Ref.~\cite{ferron04_130} 
		for the critical electric quadrupole moment ${ Q_{zz,c}^\pm = \pm 2 q_{s,c}^{ \pm } s }$.
		The internuclear distance ${ s }$ is fixed at ${ 1.6 \, {a}_{0} }$ as in Ref.~\cite{garrett08_131}; this value is 
	   close to the internuclear distance in ${ \text{CS}_{2}^{-} }$ (${ s = 1.554 \, {a}_{0} }$ \cite{NIST}).
		The corresponding  critical quadrupole moments are thus ${ Q_{zz,c}^- = -2.35152 \, e{a}_{0}^{2} }$ and ${ Q_{zz,c}^+ = 6.372016 \, e{a}_{0}^{2} }$.

\begin{figure}[htb]
			\includegraphics[width=0.90\linewidth]{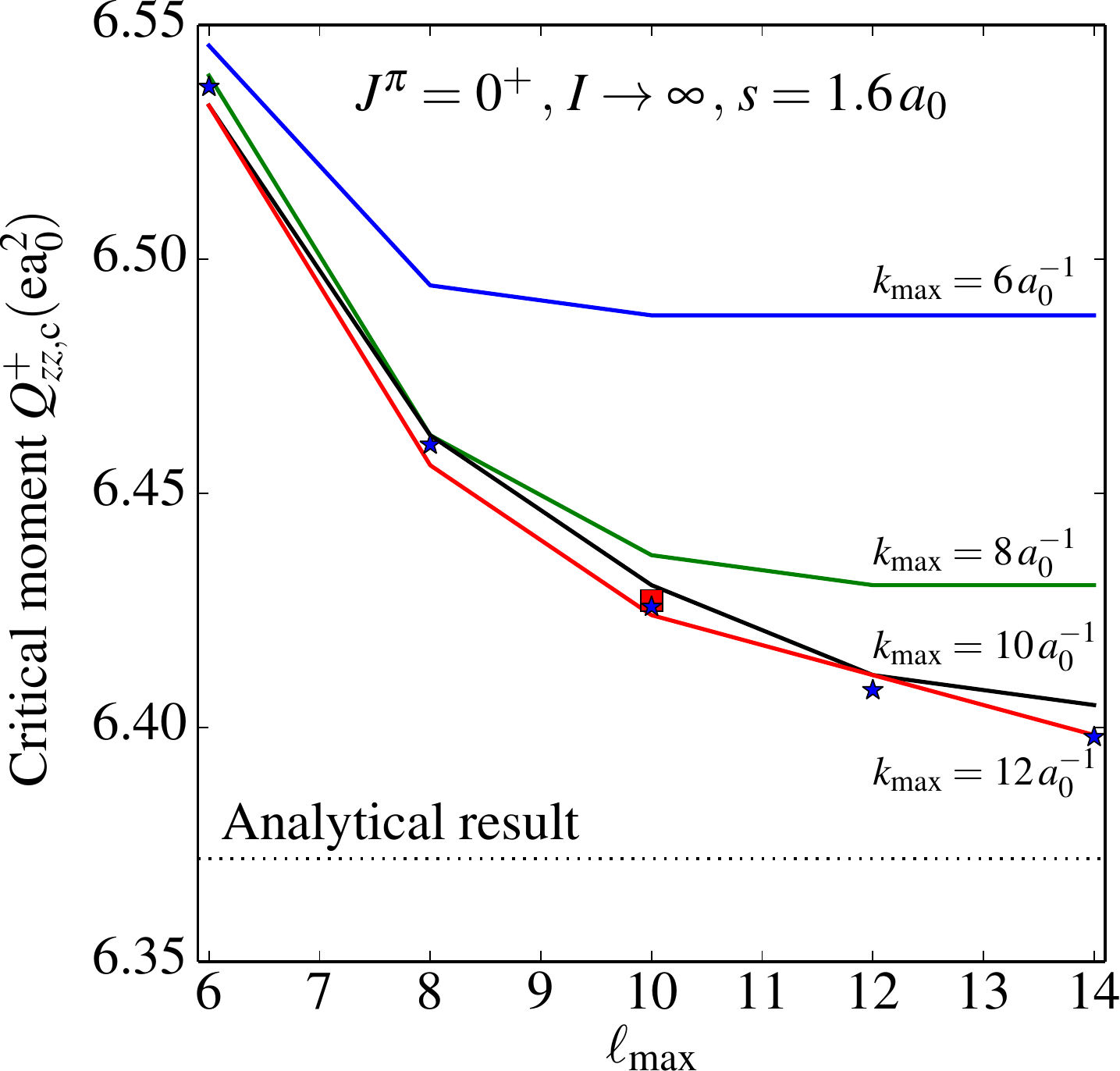}
			\caption{ Critical prolate electric quadrupole moment as a function of the orbital angular moment cutoff in coupled-channel calculations in the adiabatic limit (${ I \to \infty }$).
			The internuclear distance is fixed at ${ s = 1.6 \, {a}_{0} }$ and the corresponding value of ${ {Q}_{zz,c}^{+} = 6.372016 \, e{a}_{0}^{2} }$ is indicated by the dotted line.
			The DIM results are marked by stars. The DIM result from Ref.~\cite{garrett08_131} is denoted by a square at ${ \lmax = 10 }$.
			The convergence of the BEM results with respect to the momentum cutoff is shown for ${ \kmax = 6, 8, 10,}$ and ${ 12 \, {a}_{0}^{-1} }$.}
			\label{fig_1}
		\end{figure}

		In the DIM, the parameter that controls the accuracy of calculations is the orbital angular momentum cutoff ${ \lmax }$ that determines the size of the channel basis.
		For ${ \lmax = 12 }$, the DIM gives a critical oblate quadrupole moment of ${ {Q}_{zz,c}^{-} = -2.35162 \, e{a}_{0}^{2} }$.
		In the BEM, in addition to  ${ \lmax }$,  the momentum cutoff ${ \kmax }$ needs to be be fixed.
		By taking a real contour discretized with 80 points, 
		and ${ \kmax = 12 \, {a}_{0}^{-1} }$, 
		one obtains ${ {Q}_{zz,c}^{-} = -2.35164 \, e{a}_{0}^{2} }$.
		The  critical oblate quadrupole moment can be approached closely with both methods, 
		because it corresponds to a configuration of the attached  electron that is well localized 
		around the two positive charges at the center of the molecule.
		Thus, the  electron is expected to be primarily in low-${ \ell }$ orbits.
		For the prolate  quadrupole moment, the situation is different. Here, the attached electron, attracted by the extremal positive charges, is less  bound and
		higher-${ \ell }$ partial waves are expected to play a more important role.
		Indeed, as shown on Fig.~\ref{fig_1},
		the DIM and BEM results
		do not approach the analytical value as closely as for the oblate configuration.
		For ${ \lmax = 14 }$ (and ${ \kmax = 12 \, {a}_{0}^{-1} }$) we obtained  ${ Q_{zz,c}^+ = 6.3980 \, e{a}_{0}^{2} }$ 
		and ${ 6.3984 \, e{a}_{0}^{2} }$ with the DIM and BEM, respectively.
		While the convergence of ${ Q_{zz,c}^+ }$ with ${ \lmax }$ (and ${ \kmax }$) is slower than for ${ Q_{zz,c}^- }$,   DIM and  BEM results are fairly consistent for ${ \lmax = 14 }$ and ${ \kmax = 12 \, {a}_{0}^{-1} }$,
		and our results are in agreement with the DIM result of Ref.~\cite{garrett08_131}.

		\hspace{0.6cm}

		In realistic molecules, the effect of  Pauli blocking at short distances \cite{abdoul98_108,abdoul02_1321,garrett12_585} 
		reduces the binding in the oblate configuration; hence, in general,  
it is the prolate configuration that is more likely to bind electrons.
Thus, while the oblate configuration results are useful for  benchmarking purpose, their physical interpretation should be dealt with caution.

	\subsection{Halo scaling properties}\label{halo_scaling}

		In  quantum systems, 	the wave function may extend into the classically forbidden region to form the halo structure \cite{jensen04_233,frederico12_372}
		when the energy of a bound state  approaches the threshold.
		The halo property 
		 can appear for many types of interactions \cite{jensen04_233}.
In the case of two-body halo systems, the radial extension of the system, measured by  the root-mean-square  (r.m.s.)  radius $r^2_\text{rms} $
		changes with the separation (or detachment)  energy  ${ E }$ and $\ell$ according to the simple law \cite{riisager92_615}:
		\begin{align}\label{eq_r_rms_E_l0}
			{r}_{ \text{rms} }^{2} \propto
			\begin{cases}
			{ |E| }^{-1}  & \text{for } \ell = 0,  \\
			{ |E| }^{ - \frac{1}{2} } & \text{for } \ell = 1, 			
			\end{cases}
		\end{align}
		while for higher angular momenta ${ r^2_\text{rms} }$ stays finite when ${ |E| \to {0}^{-} }$.
		It should be remarked that deformation of the potential should not influence the halo properties \cite{misu97_1181,zhou10_1572}.
		In principle, to compare two-body halo systems at various scales, 
		the r.m.s. radius and the binding energy have to be rescaled for each potential considered.
		In the present work,  no rescaling have been applied since we do not intend to compare different halo systems.
		Our goal  to demonstrate that low-energy bound states in quadrupolar anions are behaving according to the laws expressed in Eq.~(\ref{eq_r_rms_E_l0}).

\begin{figure}[htb]
	\includegraphics[width=0.90\linewidth]{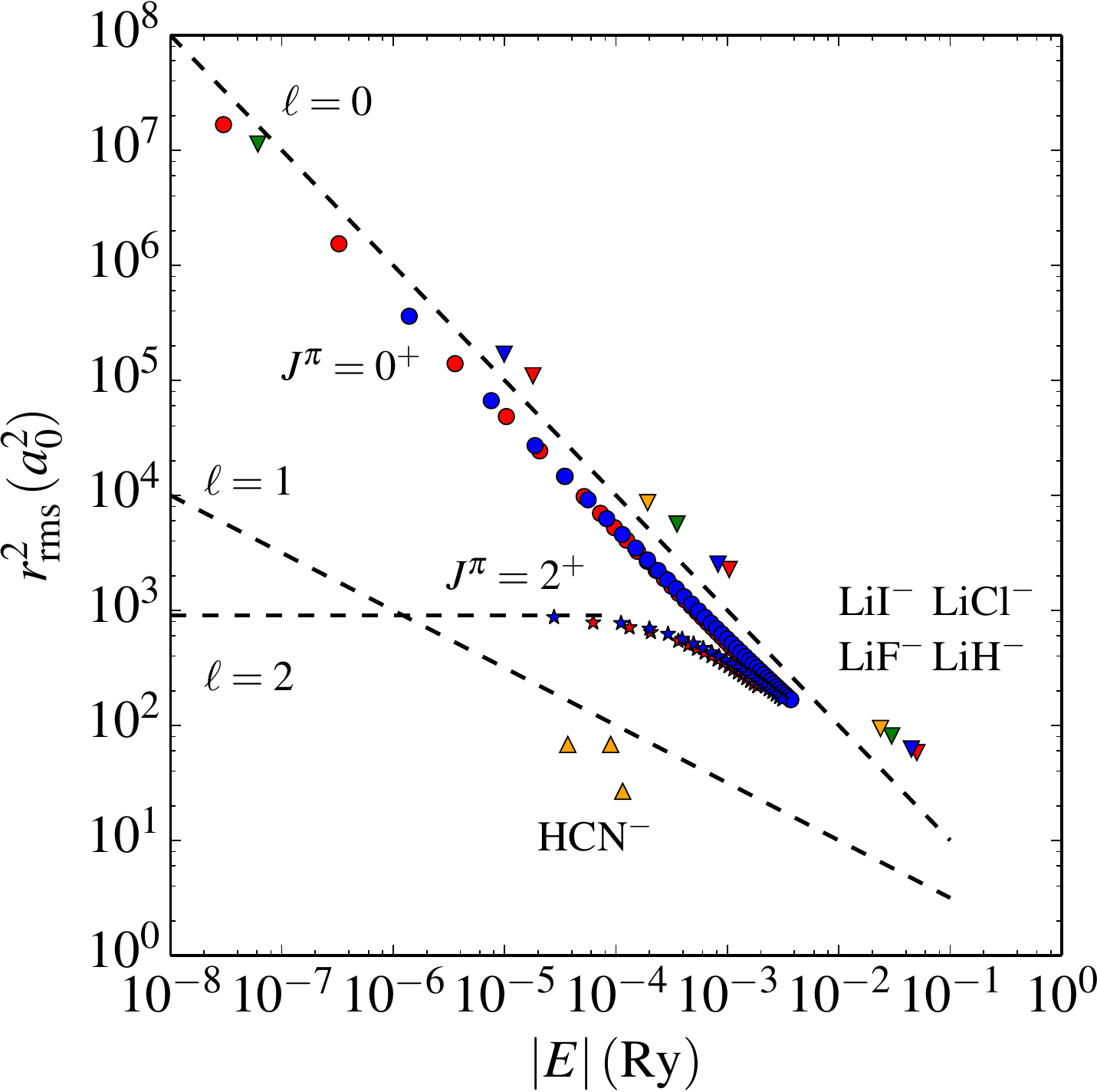}
	\caption{\label{fig_2}		
	Scaling plot $r_\text{rms}^2 (|E|)$ for quadrupolar anions as two-body halo systems. Dashed lines represent the asymptotic behavior given by Eq.~(\ref{eq_r_rms_E_l0}). The ${ \ell = 2 }$  line has been adjusted to the maximal r.m.s. radius of the ${ {J}^{ \pi } = {2}^{+} }$ bound states. Results for  ${ {J}^{ \pi } = {0}^{+} }$ and ${ {2}^{+} }$  are marked with circles and stars, respectively. Oblate/prolate states are shown in red/blue. Selected results for dipolar anions \cite{fossez13_552,fossez15_1028} are also indicated (triangles). Here,  both  $\ell = 0$ and $\ell = 1$ scaling laws are met.
}
\end{figure}

Figure~\ref{fig_2} shows $r^2_\text{rms} $ for the ${ {J}^{ \pi } = {0}^{+} }$ and ${ {2}^{+} }$ states in both oblate and prolate  configurations. The internuclear distance has been fixed at ${ s = 1.6 \, {a}_{0} }$ and the moment of inertia at ${ I = {10}^{4} \, {m}_{e}{a}_{0}^{2} }$. The quadrupole moment has been adjusted for both configurations to give a bound state at around $E \sim -1.0 \cdot {10}^{-2}$\,Ry and then gradually changed to approach the critical value. All results have been obtained using the DIM and for an orbital angular momentum cutoff of ${ \lmax = 8 }$. No bound states have been found for ${ {J}^{ \pi } = {1}^{-} }$ and ${ {3}^{-} }$.

There is no difference in the scaling behavior for oblate and prolate configurations since in both cases the bound states are dominated by the same channels. The ${ {0}^{+} }$ states have ${ \ell = 0 }$ dominant channels and  their r.m.s. radii scale according to Eq.~\eqref{eq_r_rms_E_l0}. The ${ {2}^{+} }$ states are dominated by ${ \ell = 2 }$ partial waves; here,  ${ {r}_{ \text{rms} }^{2} }$ reaches an asymptotic limit slightly below 1000\,$a_0^2$.

For completeness, selected results for dipolar anions are also shown in Fig.~\ref{fig_2} to illustrate the similarity with quadrupolar systems. The  results for the ${ {0}^{+} }$  g.s.
of ${ \text{LiI}^{-} }$, ${ \text{LiCl}^{-} }$, ${ \text{LiF}^{-} }$, and ${ \text{LiH}^{-} }$~\cite{fossez13_552}  follow the ${ \ell = 0 }$ scaling, while  the radii of ${ {0}^{+} }$, ${ {1}^{-} }$, and ${ {2}^{+} }$ states  in ${ \text{HCN}^{-} }$ \cite{fossez15_1028} ) exhibit the ${ \ell = 1 }$ asymptotic behavior. For both dipolar and quadrupolar anions, the scaling laws \eqref{eq_r_rms_E_l0} are satisfied extremely well, with ${ {r}_{ \text{rms} }^{2} }$ and ${ |E| }$ spanning about five orders of magnitude. In this sense, polar anions should be viewed as extreme halo systems.

\subsection{Rotational bands in the continuum}\label{sec_rot_band}

In a previous BEM study on dipolar anions \cite{fossez15_1028} it has been shown that the yrast band in ${ \text{HCN}^{-} }$ does not extend above the particle emission threshold. Namely, at the threshold, there appears a transition from the strong-coupling regime, in which the attached electron follows the rotational motion of the core,  to the weak-coupling regime, where the  electron's rotational motion is almost decoupled from that of the rotor.

		\begin{figure}[htb]	
			\includegraphics[width=0.90\linewidth]{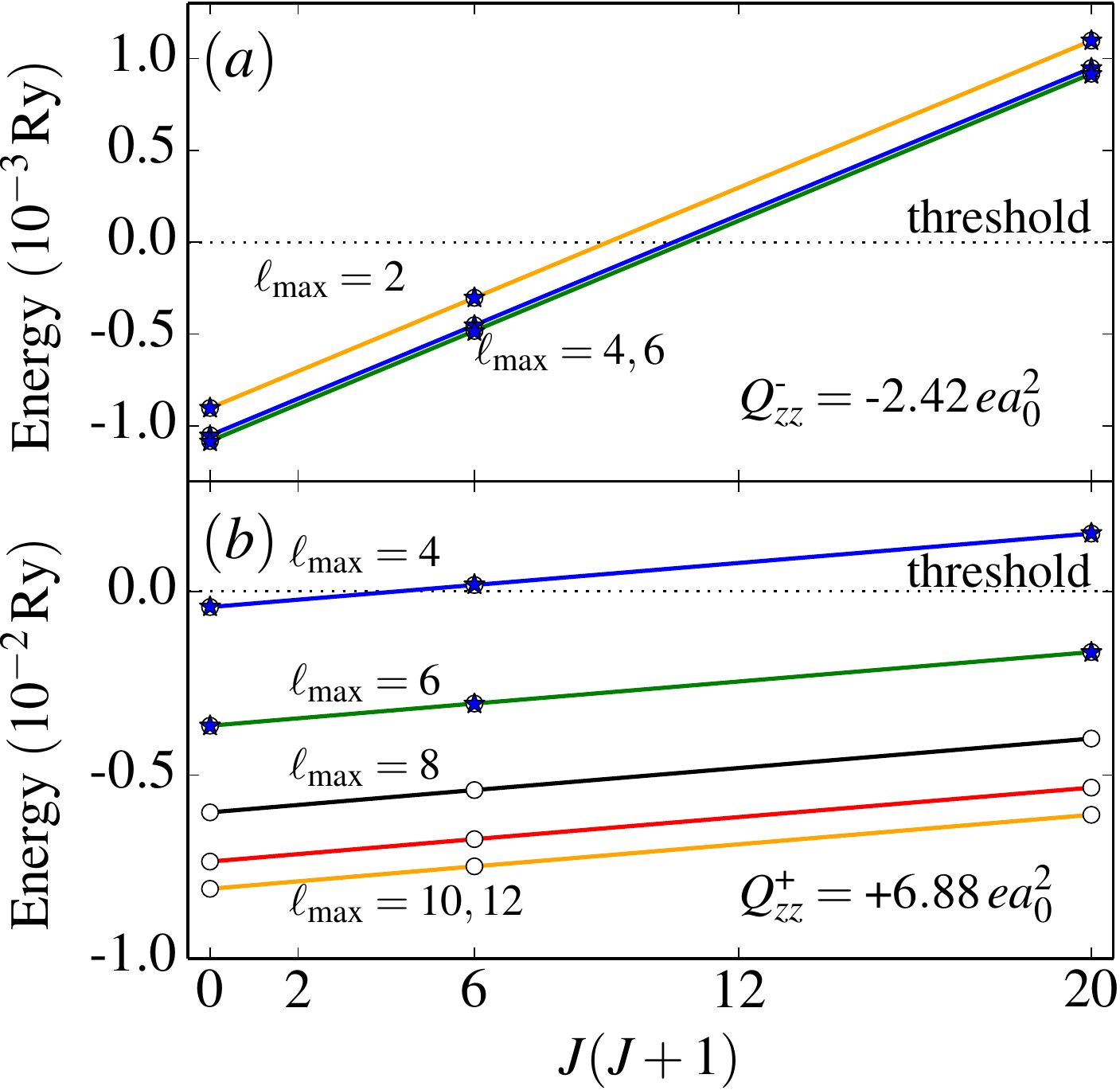}
			\caption{\label{fig_3}
			Yrast band ${ K_J = 0 }$ of quadrupolar anions defined by an internuclear distance of ${ s = 1.6 \, {a}_{0} }$, 
			a moment of inertia of ${ I = {10}^{4} \, {m}_{e}{a}_{0}^{2} }$, 
			and quadrupole moments of ${ {Q}_{zz}^{-} = -2.42 \, e{a}_{0}^{2} }$ and ${ {Q}_{zz}^{+} = +6.88 \, e{a}_{0}^{2} }$ on panels (a) and (b), respectively.
			The BEM and DIM results are denoted with empty circles and stars, respectively, 
			and are almost indistinguishable for all orbital angular momentum cutoffs considered.}
		\end{figure}
Compared to the dipolar potential, the quadrupolar potential has
		a faster asymptotic falloff ( ${ \propto 1/{r}^{3} }$) that may affect the structure of  delocalized resonant states. The impact on localized metastable states  is less obvious.
		In order to answer this question,  
		the binding energy for ${ {Q}_{zz}^{-} = -2.42 \, e{a}_{0}^{2} }$ and ${ {Q}_{zz}^{+} = +6.88 \, e{a}_{0}^{2} }$ is plotted 
	 in Figs.~\ref{fig_3}(a) and \ref{fig_3}(b), respectively, as a function of $J(J+1)$.
	 		Here we use  the same parameters as in the previous section (${ s = 1.6 \, {a}_{0} }$ and ${ I = {10}^{4} \, {m}_{e}{a}_{0}^{2} }$).
	 				The contour ${ \mathcal{L}_{c}^{+} }$ is identical for all  partial waves. It starts at zero and 
		is defined by the three points: ${ (0.3,-10^{-5}) }$, ${ (0.6,0) }$, and ${ (6,0) }$ (all in ${ {a}_{0}^{-1} }$). The three resulting segments are  discretized with 30, 30, and 40 scattering states, respectively.
		The specific values of ${Q}_{zz}$ have been chosen 
		so that the binding energy goes to zero for a total angular momentum
		$ J\approx 2,3$  at  ${ \lmax = 4 }$.
		
		The BEM and DIM results 
		 are practically indistinguishable for all the values of ${ \lmax }$ considered.
A perfect rotational behavior is predicted for both prolate and oblate configurations, even above the detachment threshold. This is 
 confirmed by the collapse of all eigenenergies to the same bandhead energy in the adiabatic limit (${ I \to \infty }$).
 At the maximal orbital angular momentum cutoff ${ \lmax }$ considered, the states in the lowest-energy (yrast) band are all dominated by the ${ \ell = 0 }$ channel at about ${ 99.7\% }$ and ${ 87.9\% }$, 
		for  the oblate and prolate configuration, respectively.
		Unlike in the dipolar case,  rotational bands of quadrupolar anions persist in the continuum.
		The widths of unbound band members  are very small (${ \Gamma \sim {10}^{-10} \, \text{Ry} }$).

\begin{figure}[htb]
			\includegraphics[width=0.90\linewidth] {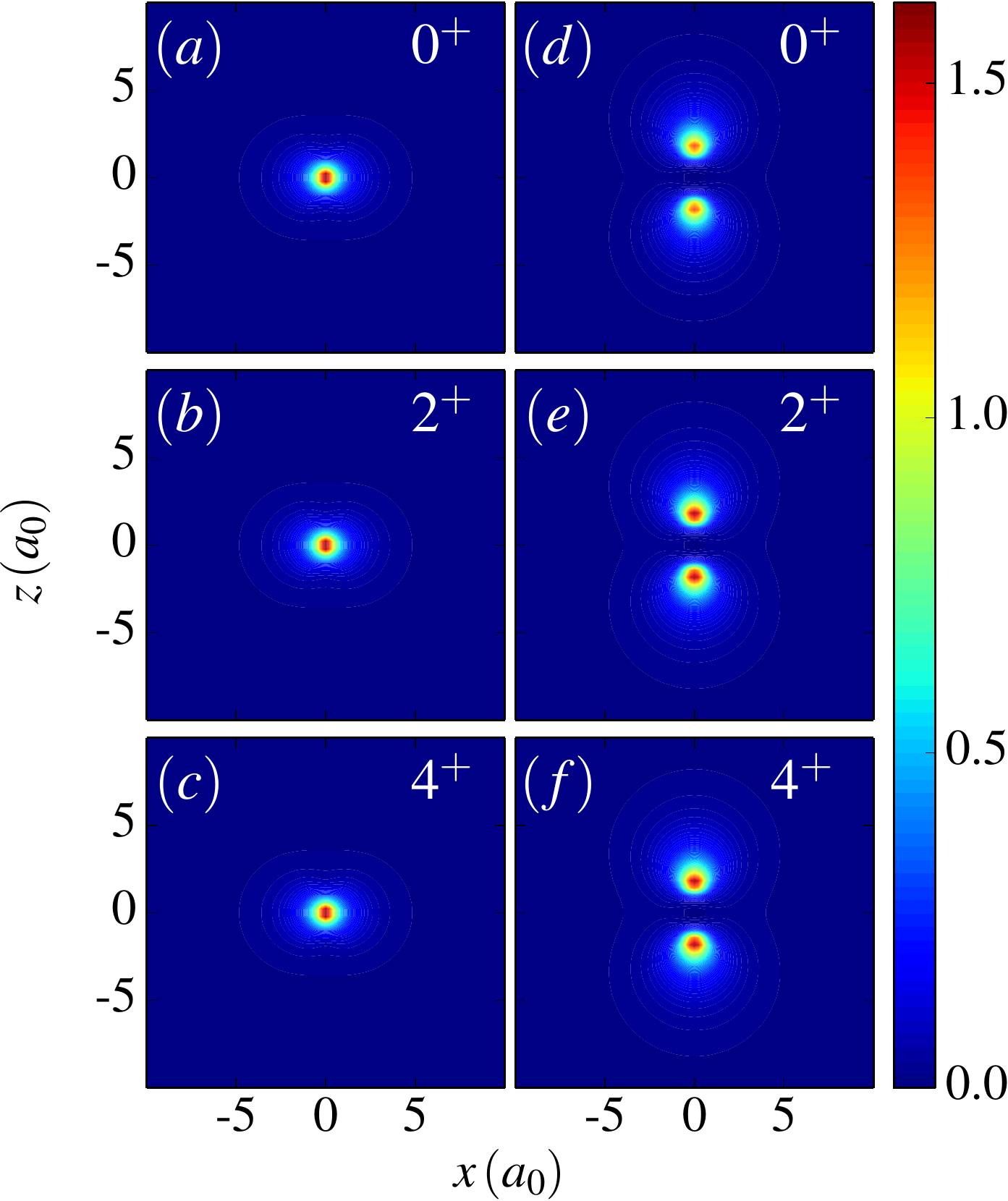}
			\caption{\label{fig_4}
			Intrinsic densities ${ \rho_{ J,0} (\vec{r}) }$ (in ${ {10}^{-2} \, {a}_{0}^{-3} }$) 
			for the yrast bands of Fig.~\ref{fig_3} calculated with ${ \lmax = 6 }$. 
			The densities for ${ {J}^{\pi} = {0}^{+} }$, ${ {2}^{+} }$, and ${ {4}^{+} }$ are shown for both oblate (a-c) and prolate (d-f) 
			 configurations.}
		\end{figure}

		In the intrinsic frame of the molecule, only the ${ K_J = 0 }$ component of the attached electron's density remains nonzero.
	Consequently, the  densities ${ { \rho }_{ J,{K}_{J} } ( \vec{r} ) }$ with ${ {K}_{J} = 0 }$  can be called intrinsic densities \cite{fossez15_1028}.
	Figure~\ref{fig_4} show the intrinsic densities for the
	 ${ {J}^{\pi} = {0}^{+} }$, ${ {2}^{+} }$, and ${ {4}^{+} }$  members 
		of oblate and prolate bands. One can see that  $\rho_{J,0}$
	 are practically identical within each band.

\subsection{Resonances}

The analysis of the unbound spectrum in quadrupolar anions can be conveniently performed using the BEM. Indeed, with the BEM one obtains the full spectrum in one diagonalization; calculations stay tractable with the increased number of channels; and the method does not require precise initial eigenvalues as in the DIM case.

Resonant spectrum calculations have been performed for the ${ {J}^{\pi} = {0}^{+} }$, ${ {1}^{-} }$ and ${ {2}^{+} }$ states in oblate and prolate   configurations (${ {Q}_{zz}^{-} = -2.42 \, e{a}_{0}^{2} }$, ${ {Q}_{zz}^{+} = +6.88 \, e{a}_{0}^{2} }$)  for ${ \lmax = 8 }$, ${ s = 1.6 \, {a}_{0} }$,  and  ${ I = {10}^{4} \, {m}_{e}{a}_{0}^{2} }$. The contour ${ \mathcal{L}_{c}^{+} }$  for each partial wave starts at zero and is defined by the three points: ${ (0.3,-10^{-5}) }$, ${ (0.6,0) }$, and ${ (12,0) }$ (all in ${ {a}_{0}^{-1} }$). The resulting segments  have been discretized with 60, 40 and 100  points representing scattering states.

\begin{figure}[htb]\includegraphics[width=0.90\linewidth]{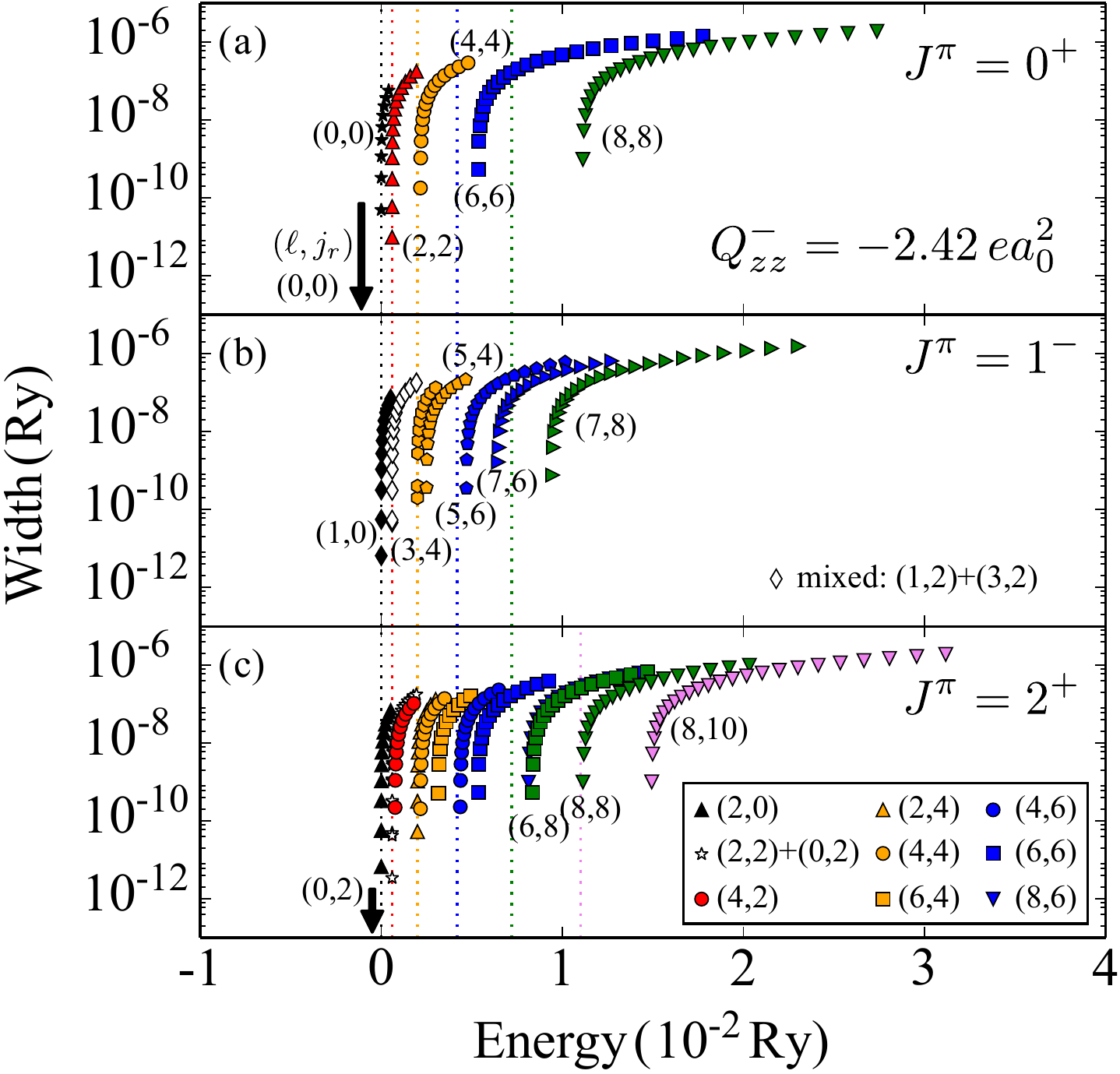}
			\caption{\label{fig_5}
			The distribution of ${ {J}^{\pi} = {0}^{+} }$, ${ {1}^{-} }$, and ${ {2}^{+} }$ resonant states in the complex energy plane for
				 the oblate configuration with ${ {Q}_{zz}^{-} = -2.42 \, e{a}_{0}^{2} }$, ${ s = 1.6 \, {a}_{0} }$, and ${ I = {10}^{4} \, {m}_{e}{a}_{0}^{2} }$ calculated with ${ \lmax = 8}$. Bound states are marked by arrows. In most cases, families of resonances are
				 characterized by one dominant channel; the corresponding labels $(\ell, {j}_{r})$ 
				 are given. Mixed groups are represented by empty symbols.
				Dashed lines indicate rotational energies of the molecule, ${ {E}_{ {j}_{r} } = { \hbar }^{2} {j}_{r} ( {j}_{r} + 1 ) / (2I) }$.
				}	
		\end{figure}

\begin{figure}[htb]
			\includegraphics[width=0.90\linewidth]{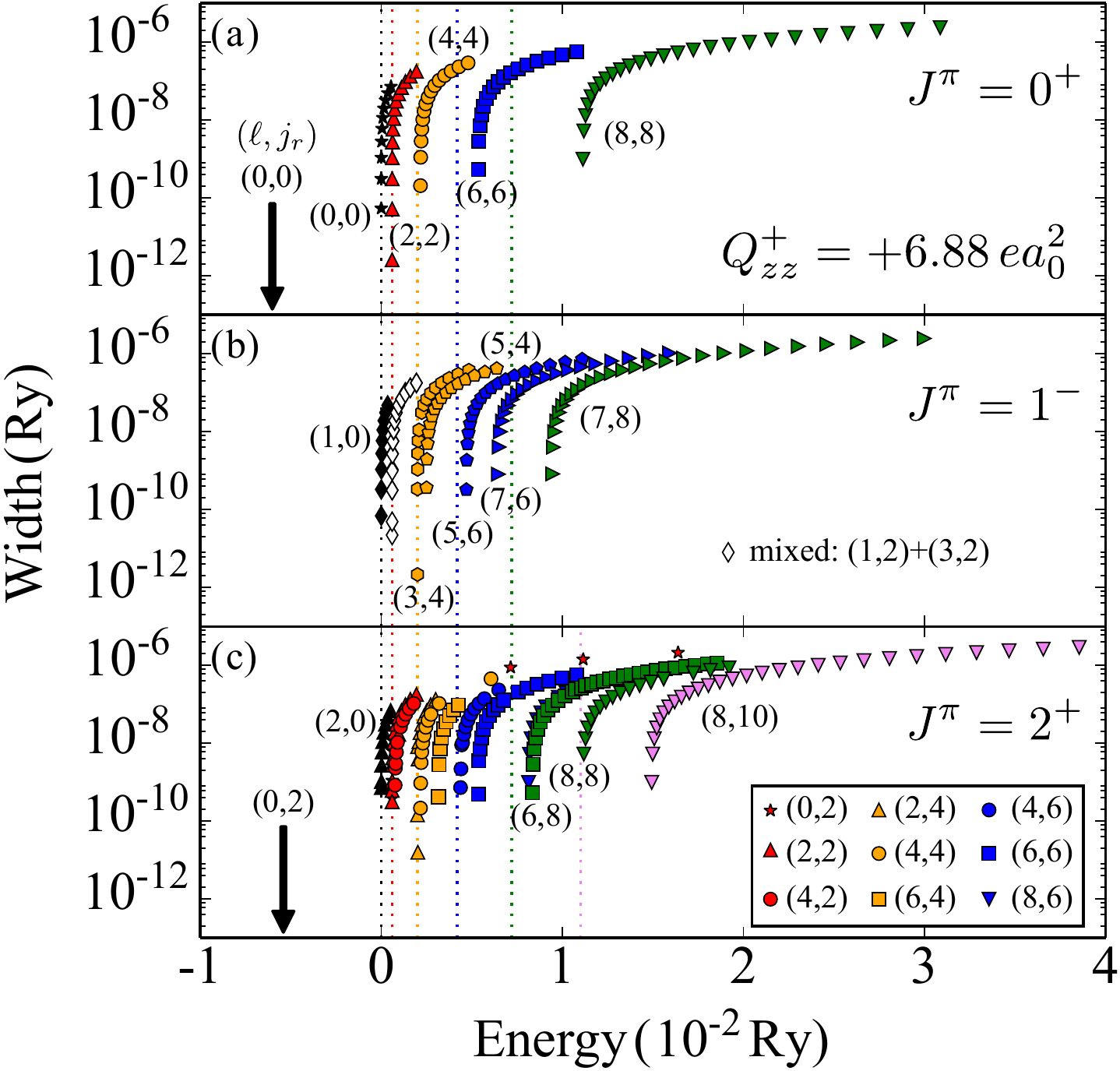}
			\caption{\label{fig_6} 
			Similar as in Fig.~\ref{fig_5} but for a prolate configuration with  ${ {Q}_{zz}^{+} = +6.88 \, e{a}_{0}^{2} }$.}
		\end{figure}

Calculations reveal the presence of families of narrow decaying resonances in the complex-energy plane as shown in Figs.~\ref{fig_5} and \ref{fig_6} for oblate and prolate configurations, respectively. The resonant structures are remarkably similar for oblate and prolate configurations, with widths ranging from  10$^{-10}$\,Ry to 10$^{-6}$\,Ry  corresponding to lifetimes in the range of ${ {10}^{-7} - {10}^{-11} \, \text{s} }$.

Each family of resonances, marked by  the same 
symbol and color in  Figs.~\ref{fig_5} and \ref{fig_6}, is characterized by one dominant channel ${ ( \ell , {j}_{r} ) }$ that represent about 99\% of the total wave function, except for those indicated by empty symbols. Within each family, narrow resonances have a very diffuse dominant-channel wave function with a small number of nodes, while broader resonances tend to have  wave functions peaked closer to the origin and having  larger numbers of nodes. Overall, energies of  resonances tend to cluster close to the rotational states of the core, except for higher excitations where significant deviations can be seen. Indeed, higher-lying resonances have larger values of the orbital angular momentum ${ \ell }$ in their dominant channel, which results in larger centrifugal barriers.

The similarity of resonant structures predicted for oblate and prolate configurations at different ${ {J}^{ \pi } }$ states can be explained in terms of the large delocalization of wave functions over thousands of ${ {a}_{0} }$. Indeed, above the particle emission threshold, even  high-$\ell$ states have the first peak of their wave function at few hundreds of ${ {a}_{0} }$. These resonances are consequently weakly sensitive to short-range details of the potential, and are mainly influenced by the asymptotic tail ${ \propto 1/{r}^{3} }$ of the quadrupolar field. Moreover,  rather small resonance widths, even for states dominated by an ${ \ell = 0 }$ channel, seem to be characteristic of multipolar  potentials since the same prediction has been made for the ${ \text{HCN}^{-} }$ dipolar anion \cite{fossez15_1028}.

The pattern of families shown in Figs.~\ref{fig_5} and \ref{fig_6} can be easily understood by considering angular momentum coupling. Namely, since for
$J^\pi=0^+$ states $\ell=j_{r}$, each family of resonances represents an electron perfectly antialigned with respect to  the rotor's angular momentum, and the the steadily increasing energy distance between groups is due to the centrifugal barrier that grows with $\ell$. The states within each family can be distinguished by their radial behavior, i.e., the number of nodes in the radial wave function.  

For $J^\pi=1^-$, the angular momentum selection rule becomes: $\ell=1$ for $j_r=0$ and 
$\ell=j_r\pm 1$ for $j_r=2, 4, \dots$. This yields 8 families (note that since
${ \lmax = 8 }$, there is only one channel with $j_r=8$). As discussed below, 
the two families with ${j}_{r} = 2$ and  $\ell = 1, 3$, marked by open symbols in Figs.~\ref{fig_5} and \ref{fig_6}, are practically degenerate. This explains the multiplicity of families for $J^\pi=1^-$, and -- in a similar way -- for $J^\pi=2^+$.

Because many resonances belonging to low-$\ell$ channels cluster around  the rotational states of the molecule, the density of resonances in the complex energy plane is high, and accidental (near-)degeneracies occur. This results in a strong configuration mixing. Such
families of resonances have two dominant channel wave functions at low energy, and are referred to as ``mixed'' groups in the following. 
In Figs.~\ref{fig_5} and \ref{fig_6}, mixed groups are
the (1,2) and (3,2) families for $J^\pi=1^-$ and
the (0,2) and (2,2) families for $J^\pi=2^+$.

\begin{figure}[htb]
			\includegraphics[width=0.90\linewidth]{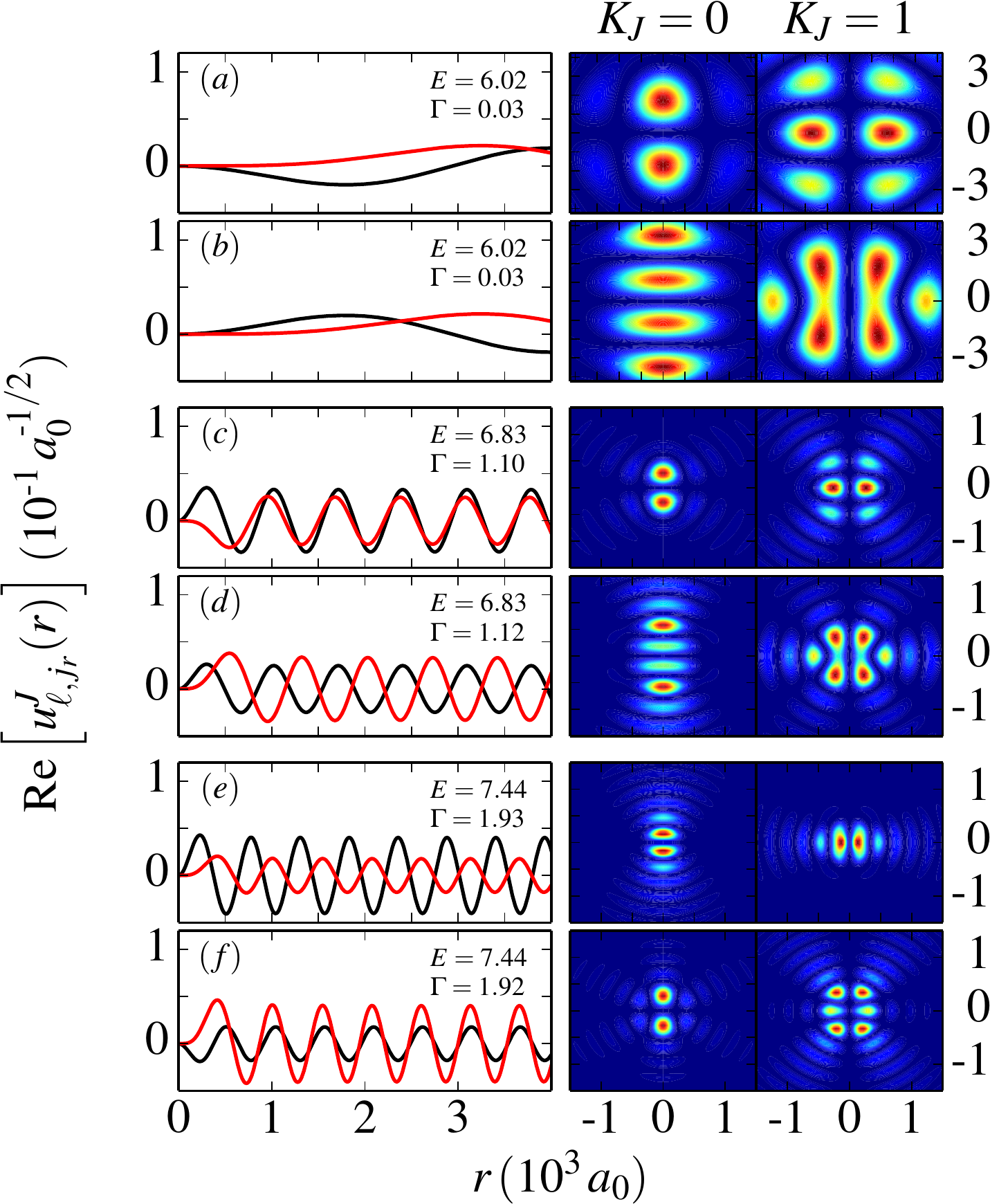}
			\caption{\label{fig_7} Left: Real parts of the two dominant channel wave functions of selected quasi-degenerate ${ {J}^{ \pi } = {1}^{-} }$ resonances of  Fig.~\ref{fig_6}(b). 
			The imaginary parts of the channel wave functions are about four orders of magnitude smaller. Resonance energies (in $10^{-4}$ Ry)
			and widths(in $10^{-8}$ Ry) are indicated. The channel wave functions (1,2) and (3,2) are colored  black and red, respectively.
				Right: Corresponding intrinsic densities (in arbitrary units) for ${ {K}_{J} = 0 }$ and 1.
				}	
		\end{figure}

Within each of these groups, there appear pairs of resonances, or ``doublets'', with very close complex energies. To illustrate the strong mixing between overlapping resonances, in Fig.~\ref{fig_7} we show
the wave functions of six ${ {J}^{ \pi } = {1}^{-} }$ doublets, belonging to  the mixed group at the prolate configuration of Fig.~\ref{fig_6}(b).
Their dominant channel wave functions are 
$(\ell = 1, j_r = 2)$ and $(\ell = 3, j_r = 2)$. 
		For the lowest-energy doublets in a mixed group the channel mixing is maximal: the channel wave functions  are almost identical and the total wave functions can be represented by  their symmetric and asymmetric combinations, as in a textbook case of a two-state mixing.
		However, as the excitation energy increases, the doublets move apart slightly in the complex energy plane and the channel wave functions start to differ. However, the configuration mixing still remains strong.
Moreover, as one can see in Fig.~\ref{fig_7}, while the intrinsic densities
		for ${ {K}_{J} = 0 }$ and 1 are very different within a given doublet, 
		they show similar structures for states within the same mixed group.
		At low energies,  the ${ {K}_{J} = 0 }$ term dominates  (with a weight of 71\%) in each doublet
		and then its weight increases to about 74\% for one state and decreases to about 68\% for the other state of the doublet.

	\subsection{Evolution of the spectrum with the quadrupole moment}

 As demonstrated above,
for the supercritical quadrupolar molecules with ${ | Q_{zz}^{ \pm } | > | Q_{zz,c}^{ \pm } | }$, many resonances can exist in vicinity of   the rotor energies. The subcritical quadrupolar molecules with ${ | Q_{zz}^{ \pm } | < | Q_{zz,c}^{ \pm } | }$ may still accommodate  resonances in spite of their weaker quadrupolar field.

\begin{figure}[htb]
	\includegraphics[width=0.90\linewidth]{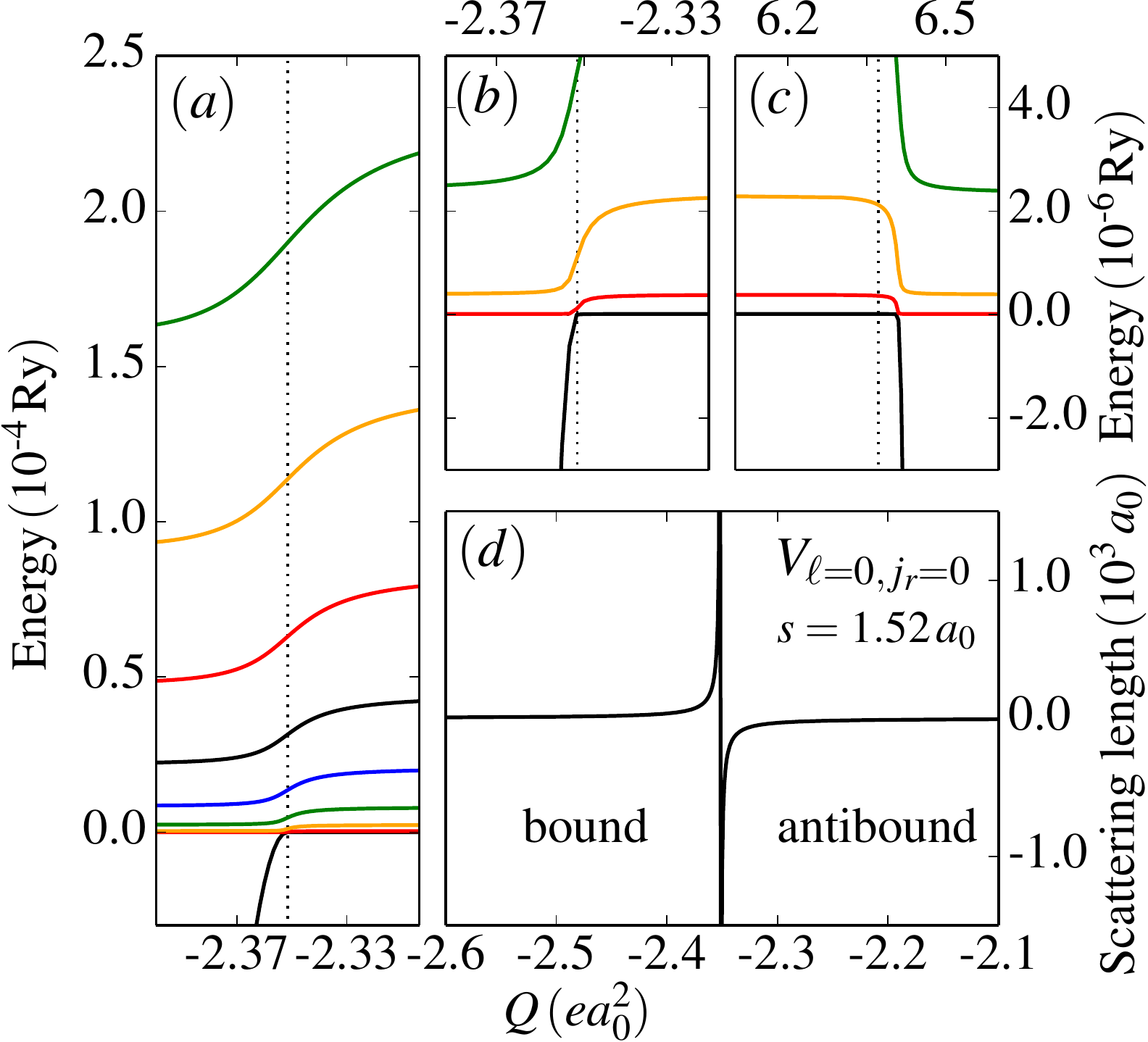} 
	\caption{\label{fig_8} The low-lying ${ {J}^{ \pi } = {0}^{+} }$ eigenenergies (real parts) of a quadrupolar anion as a function of the electric quadrupole moment in the vicinity of $Q_{zz,c}^{-}$ (a,b) and  $Q_{zz,c}^{+}$ (c).
	Panel (d) shows the scattering length of a  scattering state at $E= {10}^{-12}$\,Ry, which an eigenstate of the diagonal channel-channel coupling potential ${ {V}_{ c , c } }$ with ${ c = ( \ell = 0 , {j}_{r} = 0 ) }$ for ${ s = 1.52 \, {a}_{0} }$.
}
\end{figure}

The transition from a supercritical to subcritical quadrupolar anion is illustrated in Fig.~\ref{fig_8} for ${ {J}^{ \pi } = {0}^{+} }$ states. Figure~\ref{fig_8}(a) shows real energies of the lowest resonant states  as a function of the quadrupole moment in an oblate system. If one denotes the energy of the ${ i }$-th resonance of a supercritical molecule outside the critical region  as ${ {E}_{i} }$, then by changing the electric quadrupole moment continuously beyond  $Q_{zz,c}^{-}$, one arrives at ${ {E}_{i} \to {E}_{i}' \approx {E}_{ i + 1 } }$. A close look at the area in the immediate vicinity of the critical quadrupole moment in Figs.~\ref{fig_8}(b) ($Q_{zz,c}^{-}$)
and \ref{fig_8}(c) ($Q_{zz,c}^{+}$) one can see that  this rearrangement of eigenvalues happens at the critical values.
Moreover, for ${ | Q_{zz}^{ \pm } | << | Q_{zz,c}^{ \pm } | }$ eigenenergies are almost equal for oblate and prolate  configurations, as the corresponding wave functions are hardly sensitive to details of the potential.

Such an eigenvalue rearrangement  at the critical quadrupole moment suggests a critical behavior of the system \cite{lassaut96_1467}. The details can be understood by looking at the behavior of the g.s. eigenenergy at the threshold \cite{moiseyev98_92}. Indeed as shown in Ref.~\cite{klaus80_1490}, the bound state energy approaches zero according to:
\begin{equation}
			{E}_{ \sigma \to { \sigma }_{c}^{+} } \sim { ( \sigma - { \sigma }_{c} ) }^{ \alpha },
			\label{eq_low_E_critical_behavior}
\end{equation}
		where ${ \sigma = {Q}_{zz} }$, ${ { \sigma }_{c} = {Q}_{zz,c} }$, and ${ \alpha }$ is the critical exponent.
For a spherical potential with an asymptotic falloff ${ \propto 1/{r}^{3} }$, 
		the critical exponent should be  two \cite{klaus80_1490,lassaut96_1467}. In the case considered,
		 the quadrupolar potential is not isotropic, but the dominant g.s. channel wave function has ${ \ell = 0 }$.

A fit of the bound-state energy  for the oblate configuration,  in the range of ${ {Q}_{zz}^{-} \in [ -2.4 , {Q}_{zz,c}^{-} ] }$, yields ${ {Q}_{zz,c}^{-} = -2.35175 \, e{a}_{0}^{2} }$ and ${ \alpha = 2.00006 }$, in agreement with analytical results. The transition of a bound state dominated by ${ \ell = 0 }$ to a resonance in quadrupolar anions appears to be a second-order  transition \cite{serra01_259}, which is understood as a continuous change of the system.
This means  that there is no bound state at the threshold  \cite{klaus80_1490}.
The shift of all resonance energies at the critical quadrupole moment is a consequence of this transition, with the bound state changing continuously to reach the energy of the first resonance state. This results in an avoided crossing between the g.s. and the first excited state, which then propagates to all excited states. Avoided crossings in the eigenvalue spectrum of the Hamiltonian reveal the existence of exceptional points \cite{okolowicz03_21,hernandez11_498,heiss12_1392} in its complex extension.

The criticality  can also be assessed by considering the scattering length ${ a }$ of the system at different values of ${Q}_{zz}$. At low-energy (${ k \to 0 }$), the scattering length is related to the ${ \ell = 0 }$ phase shift ${ { \delta }_{0} (k) }$ of a scattering state through:
		\begin{equation}
			\lim_{ k \to 0 } \frac{k}{ \tan \delta (k) } = -\frac{1}{a_0}.
			\label{eq_minus_inv_scatt_length_s_wave}
		\end{equation}
In our calculations, all partial waves are included up to a given orbital angular moment cutoff ${ \lmax }$, and even if the ${ \ell = 0 }$ component dominates at low energy, there  still exist small contributions coming from higher partial waves. In order to illustrate the criticality of the system in a simple case, only the ${ \ell = 0 }$ diagonal element of the channel-channel coupling potential has been considered. Indeed,  the scaled parameter ${ q_{s,c} = qs }$ reaches a constant~\cite{ferron04_130} at the critical value. Therefore, by changing ${ s \to s' }$ so that the ${ \ell = 0 }$ diagonal element of the potential becomes more and more important, one can effectively evolve the negative critical quadrupole moment ${ Q_{zz,c,0}^{-} = - 2 q_{s',c} s' }$ for ${ \ell  = 0 }$ to the  value $Q_{zz,c}^{-}$ of the complete problem. To this end, one has to decrease ${ s }$, or conversely  increase ${ {q}_{s,c} }$, to localize the electron in a (almost) pure ${ \ell = 0 }$ bound state.
		
In practice, by considering only $\ell=0$ wave, one obtains ${ s = 1.52 \, {a}_{0} }$ instead of ${ 1.6 \, {a}_{0} }$ for the full problem. The scattering length is plotted as a function of $Q_{zz,c,0}^{-}$ in Fig.~\ref{fig_8}(d), and shows a characteristic divergence at the critical value. 
Such behavior corresponds to the formation of the Feshbach resonance. In the pure ${ \ell = 0 }$ case, the negative scattering length is associated with a virtual state, but in the full problem higher partial waves prevent formation of  a resonance as shown in of Fig.~\ref{fig_8}(b).
This observation on a transition of a bound state into the continuum, together with the divergence of the scattering length, are general features of open quantum systems.  

\section{Conclusion} \label{conclusion}

In this work, we studied  bound and unbound states in quadrupolar anions in a nonadiabatic molecule-plus-electron picture.  The Schr{\"o}dinger equation of the system, expressed in a  coupled-channel form,  was solved by a diagonalization in the Berggren basis   or by means of a  direct integration. The  BEM and the DIM approaches have been benchmarked  against analytical results for the critical electric quadrupole moment. It is shown that  binding energies and r.m.s. radii of bound states in  oblate and prolate configurations of a quadrupolar anion follow the two-body halo scaling properties over several orders of magnitude. Using the density of the attached electron in the molecular frame, as well as the collapse of g.s. eigenenergies to the bandhead energy in the adiabatic limit, we demonstrated the existence of regular rotational  bands   below and above the detachment threshold.   
The presence of the strong coupling of electron's motion to the molecular core above the threshold makes the situation in quadrupolar anions different from that in  dipolar anions, where
electron's motion in a resonance state becomes largely decoupled from molecular rotation \cite{fossez15_1028}.

We demonstrated the presence of families of narrow resonances  close to the rotational states of the molecule. The unbound spectrum contains  many quasi-degenerate states, forming  regular rotational bands. The presence of narrow resonances close to the threshold, even for subcritical values of  $Q_{zz}$, may produce a low-energy peak in the cross section. Finally, the evolution of a bound state into the continuum can described in terms of a second order transition with the critical exponent $\alpha=2$. 

In summary, this work shows that quadrupolar anions are spectacular realizations of open quantum systems. They exhibit fascinating behavior around the detachment threshold, such as halo structures, overlapping resonances, Feshbach resonances, and critical behavior. Consequently, these simple polar molecules constitute an ideal laboratory of weakly bound and unbound quantum states.

\begin{acknowledgments}
This material is
based upon work supported by the U.S.\ Department of Energy, Office of
Science, Office of Nuclear Physics under  award numbers 
DE-SC0013365 (Michigan State University) and
DE-FG02-10ER41700 (French-U.S. Theory Institute for Physics with Exotic Nuclei).
\end{acknowledgments}


%

\end{document}